\documentclass[prb,superscriptaddress,twocolumn,showpacs]{revtex4-1} 
\usepackage{epsfig}
\usepackage{epstopdf}
\usepackage{multirow}
\usepackage{times}
\usepackage{dcolumn}
\usepackage{graphicx}
\usepackage{ulem}
\usepackage{color}

\begin{document}
\title{Surface antiferromagnetism and incipient metal-insulator transition in strained manganite films}
\author{F. Cossu}\author{U. Schwingenschl\"ogl}\affiliation{Materials Science and Engineering, KAUST, 23955-6900 Thuwal, Kingdom of Saudi Arabia}
\author{G. Colizzi}\author{A. Filippetti}
 \author{Vincenzo Fiorentini}
\affiliation{CNR-IOM and Dipartimento
di Fisica, Universit\`a di Cagliari, I-09042 Monserrato (Ca), Italy}

\date{\today}

\begin{abstract}
Using  first-principles calculations,  we show that the (001) surface of the ferromagnet La$_{0.7}$Sr$_{0.3}$MnO$_{3}$ under an epitaxial compressive strain favors antiferromagnetic order  in the surface layers, coexisting with ferromagnetic bulk order. Surface antiferromagnetism  is accompanied by a  very marked surface-related spectral pseudogap, signaling an incomplete metal-insulator transition at the surface. The different relaxation and rumpling of the  MnO$_{2}$ and LaO surface planes in the two competing magnetic phases cause distinct work-function changes, which that are of potential diagnostic use. The AF phase  is recognized as an extreme surface-assisted case  of the combination of in-plane AF super-exchange and vertical FM double-exchange couplings that rules magnetism in manganites under in-plane compression.
\end{abstract}

\maketitle

 Rare-earth manganites are  promising
 for applications to magnetic sensors,\cite{jin-JAP1994,balcells-APL1996} spintronic
devices,\cite{pallecchi-PRB2005,
yin-APL2000,hueso-Nature2007,liu-ML2008,mathews-Science1997,millis-Nature1998}
and  fuel-cell components,\cite{lussier-TSF2008,miyazaki-jPS2008} and have been the focus of many theoretical studies.\cite{tokura-Science2000,yunoki1-PRL1998,colizzi-PRB07,colizzi-PRB08}
The colossal-magnetoresistant above-room-temperature ferromagnet (FM) La$_{1-x}$Sr$_{x}$MnO$_{3}$ (LSMO henceforth) around $x$=1/3 is one of the most studied  in this class. Its magnetic phase and structural properties are quite sensitive to hydrostatic as well as uniaxial strains.\cite{colizzi-PRB07,colizzi-PRB08} It is natural to expect that such effects will be amplified  at surfaces and that tight magnetic phase competition may ensue,  as suggested by previous theoretical work \cite{a1,a2,pruneda07} and by experimental observations of a lowered magnetic moment in substrate-constrained LSMO films.\cite{tebano-PRB06,lehmann-EPJB2007}
Here we demonstrate, using ab initio simulations, that indeed antiferromagnetism (AF) takes over at the (001) surface of LSMO under appropriate in-plane compression, and it coexists   with  FM order inside the bulk. The surface exhibits large  rumpling relaxations which affect the work function, and its electronic properties suggest an incipient surface metal-insulator transition and reduced in-plane conductivity. 

%
First-principles calculations are performed using the PAW method \cite{paw} within the generalized gradient approximation to  density functional theory  with Hubbard correction \cite{anisimov} (GGA+U) in the  Dudarev  formulation\cite{dudarev} as implemented in the VASP code.\cite{vasp}  The only independent  parameter, U--J, is set   to 2 eV as in our previous work on bulk LSMO.\cite{colizzi-PRB07,colizzi-PRB08} GGA+U opens a gap at the Fermi level (E$_{F}$) in the minority channel,\cite{colizzi-PRB07,colizzi-PRB08,ma} in agreement with the known semimetallic character of LSMO. In the plain gradient approximation (GGA), E$_{F}$ barely slices through the lower 
minority conduction band,
so the GGA  structure and magnetism are in fact largely similar to GGA+U, and are not discussed further here. The energy cutoff is 400 eV, and  k-point grids up to  9$\times$9$\times$2 are used.

We simulate the epitaxially-constrained Mn-terminated (001) LSMO surface with a 13-layer symmetric slab with  $\sqrt{2}$$\times$$\sqrt{2}$ (two cations per plane) in-plane section, and containing  seven MnO$_2$ and six (La/Sr)O layers at different in-plane strains.
Doping at $x$=1/3 is effected by Sr substitution of one La per LaO  layer,  symmetrically with respect to
the central MnO$_{2}$ layer;  the subsurface LaO layers are Sr-free. The periodic slab image are separated by $\simeq$10 \AA\, of vacuum. All ionic degrees of freedom are relaxed starting from
 the bulk rhombohedral $R$3$c$ structure, with force threshold 0.01 eV/\AA.

\begin{figure}[h]
\centering
\includegraphics[width=6cm]{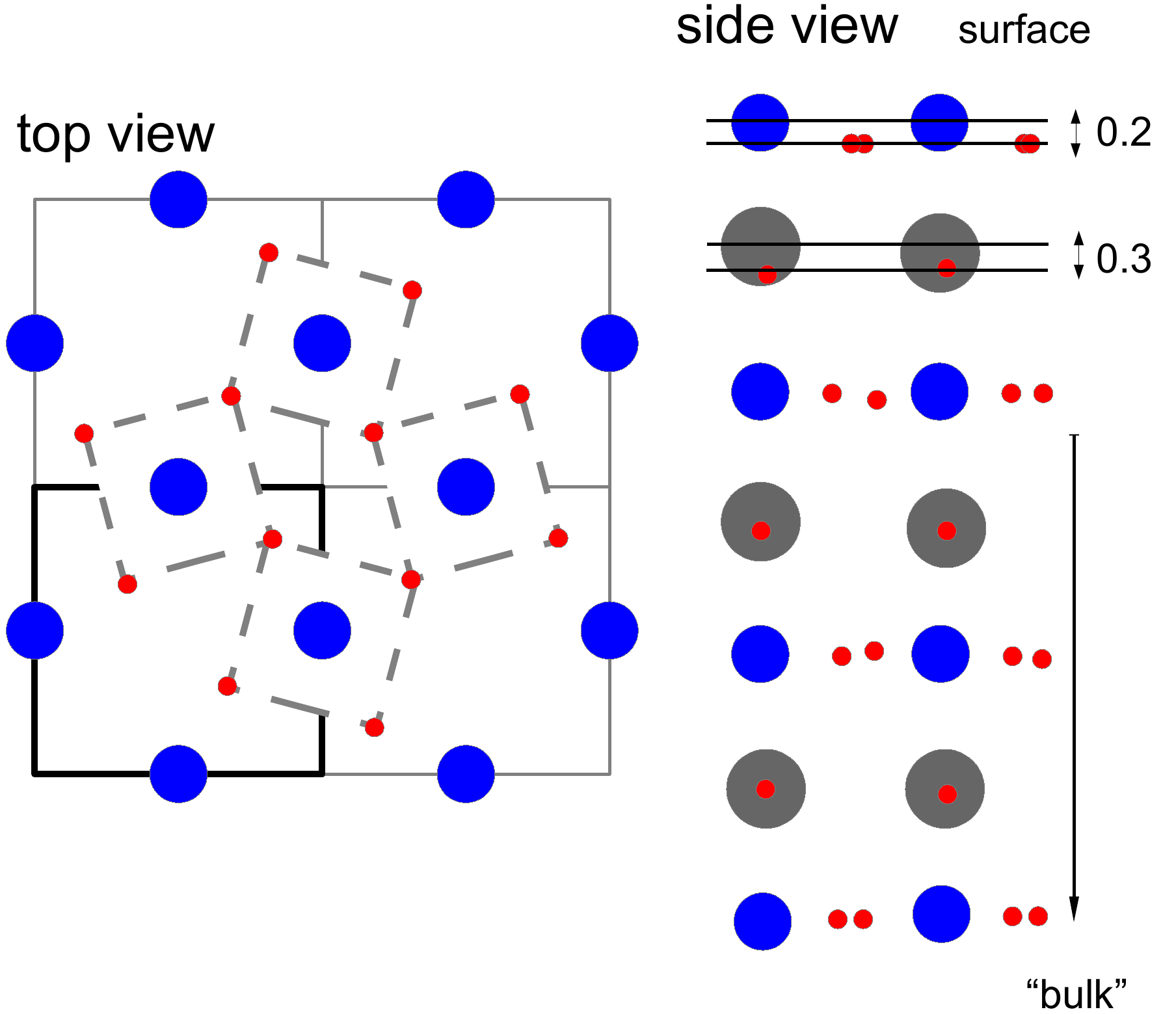}
\includegraphics[height=4.8cm]{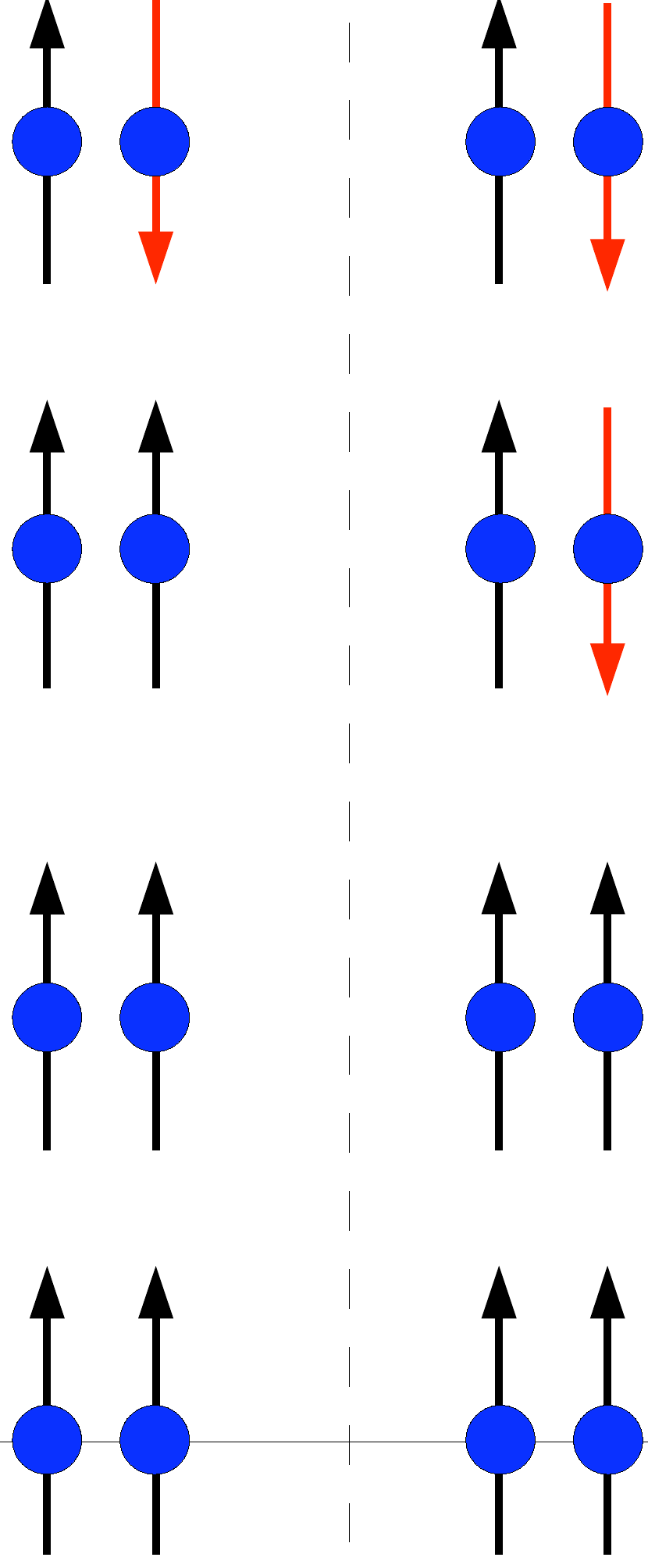}
\caption{(Color online) Sketch of the simulation supercell (small red circles: O;  large gray circles: La/Sr; intermediate blue circles: Mn). Left: top view (in-plane surface cell outlined). Center:  side view (including vertical rumpling of surface layers). Right: 1f and 2f magnetic structures (surface is upper side).
\label{slab-sketch}}
\end{figure}

 With a fairly extensive search for competing surface magnetic structures we restricted the comparison  to the FM, {\bf q}=(0,0,0), and to structures we call ``1f'' and ``2f'', namely FM phases with Mn moments in the surface and, respectively, subsurface MnO$_2$  layer being spin-flipped according to an  AF$_{C}$ pattern with {\bf q}=(1/2,1/2,0) as in Fig.\ref{slab-sketch}, rightmost panel. Magnetic patterns which are not energy-competitive are the same as 1f or 2f with three or more Mn layers flipped, and others locally mimicking bulk AF$_A$ ({\bf q}=(0,0,1/2)) and AF$_G$ ({\bf q}=(1/2,1/2,1/2)). The magnetic behavior is found to be robust with respect to slab thickness (from 7 layers to the 13 layers  used here).

\begin{figure}[h]
\includegraphics[width=7.5cm]{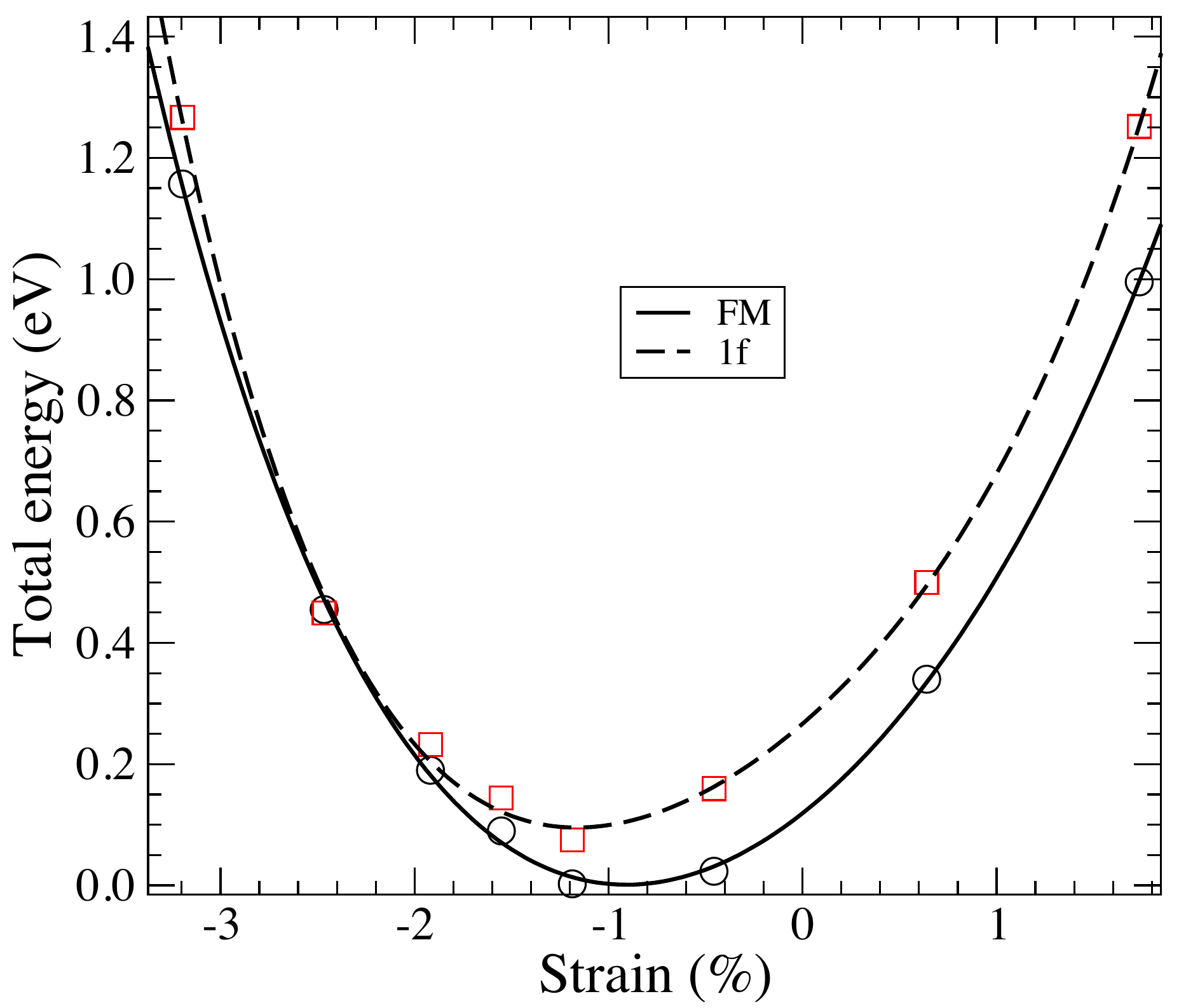}
\caption{(Color online) Calculated energies vs strain for the FM (circles) and 1f (squares) structures. Lines are fourth-order polynomial fits; the energy zero is the minimum of the FM fit.
\label{totener}}
\end{figure}

The energies of 1f and  FM vs strain are shown in Fig.\ref{totener}. Zero strain corresponds to  the experimental bulk lattice constant of LSMO (3.871 \AA). The results indicate a stability region for 1f at strains around --2.5\%. At larger compressive strains the FM phase becomes favored again (although more complex AF phases\cite{tebano-PRB06} not considered here may start playing a role), whereas for tensile strains no competition occurs: the FM is always favored. 
 Fig.\ref{totener} also shows that in both phases  the  surface at zero strain  is subject to a tensile surface stress. A rough estimate from the linear term of the energy\cite{marcus} around zero strain gives a surface stress of 0.20 eV/\AA$^{2}$ which is comparable to the very large\cite{pt100} 0.35 eV/\AA$^{2}$ of  Pt (001). The finite-thickness simulation slab is  contracted by the surface stress, and is in equilibrium at a non-zero compressive strain,  which indeed corresponds to the release of a tensile stress.  Thus, LSMO thin films will adapt more easily to shorter-lattice constant substrates, less so to larger-lattice constant ones. (Specifically, at the present film thickness of about 1 nm the effective strain needed to stabilize the 1f structure is about --1.5\%.)

The raw total energies of 1f and FM (Fig.\ref{totener}) show that in the vicinity of --2.5\% strain  the  first surface layer of LSMO (001) will be antiferromagnetic. Below we  analyze the structure and magnetic properties at this strain, which, we note, corresponds to growth on LaAlO$_{3}$.  While the 1f (top-layer AF) phase is favored, the 2f phase is disfavored, indicating that AF will hardly penetrate deeper than one Mn layer into the bulk. 
Even  if we factor out the bulk energy using  a surface-specific  energy gain
$ \Delta \gamma$=$(\Delta E_{s}$--$n$$\Delta E_{b}$)/${2A}$
in lieu of the raw energy difference ($\Delta E$$_{s}$ is the energy difference  between  the FM slab and the spin-flipped slab, $\Delta$$E$$_{b}$ is twice --there being two Mn per layer-- the bulk energy difference per formula unit between AF$_{C}$ and FM, $n$  the number of spin-flipped layers, $A$=28.6 \AA$^{2}$ the surface area), the 2f structure still gains zero energy to numerical accuracy, and all other structures give negative  gains.  This   confirms that only the topmost layer of the surface region of LSMO (001) is likely to exhibit antiferromagnetism. Indeed, AF is {\it expected} to be surface-induced and surface-confined: a previous analysis  \cite{colizzi-EPJ}  of bulk LSMO under strain has shown  that the vertical magnetic couplings are large and positive (i.e. FM) for all reasonable epitaxial conditions and external strains. 
A corollary, confirmed by the calculation, is that larger strains will not help AF to dive any deeper into the bulk.

\begin{figure}[ht]
\includegraphics[width=8cm]{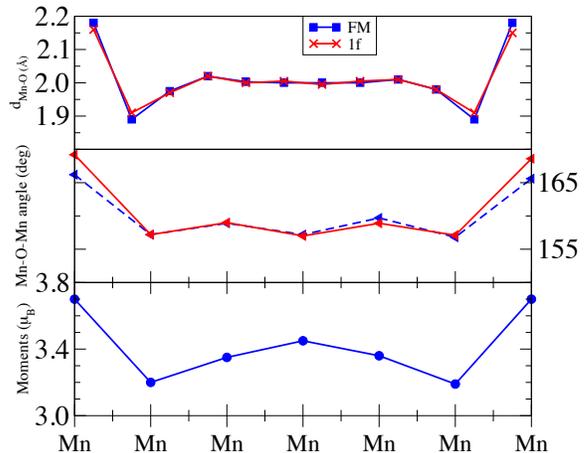}
\caption{(Color online) Structural parameters vs position (top surface: left; bottom  surface: right). Top: Mn-apical O interlayer distance; center: octahedron rotation arouns $\hat{s}$; bottom, Mn magnetic moments.
 \label{Udata}}
\end{figure}

  The strained surface exhibits quite outstanding  features for a metallic surface: i) a very large outward  relaxation of the top layer (almost 20\% compared to bulk interlayer spacing), associated with ii) an increased (+10\%) Mn magnetic moment at the surface, and iii) a large rumpling (over 10\% on average) in the top MnO$_{2}$ and LaO layers. A selection of calculated quantities is  in Fig.\ref{Udata}. Mn-O octahedra behave nearly as rigid units, except in the surface and subsurface layers, where structural and magnetic behavior change in lockstep. 
The tilt angles of the octahedra off the surface normal $\hat{s}$$\equiv$(001) are   essentially zero all over the slab, i.e. octahedra are  perfectly aligned along $\hat{s}$ and laterally rigid. The large rotations of the octahedra around $\hat{s}$ are roughly constant in the bulk region, with a Mn-O-Mn angle $\alpha$=160$^{\circ}$ as in bulk LSMO at the same strain;   the rotation then increases  in the subsurface layer ($\alpha$=155$^{\circ}$), and finally decreases in the surface layer ($\alpha$=170$^{\circ}$). Concurrently, 
the vertical (i.e. along $\hat{s}$) distance between Mn  in the top layer and O in the LaO layer just below increases to 2.2 \AA\, (+10\%  relative to the bulk 2.0 \AA), while that between the subsurface Mn and O in the LaO layer below  drops to 1.9 \AA\, (--5\%). This correlates with the Mn magnetic moment, which
drops from 3.4 $\mu_{B}$ in the bulk to 3.3  $\mu_{B}$  in the subsurface layer (--6\%), and then increases to 3.7  $\mu_{B}$ in the top layer  (+9\%). 
\begin{table}[ht]
\caption{Rumpling  (\AA) of the top surface layers.} \label{off-centering_table}
\centering \begin{tabular}{lccc}              & {FM} &  {AF} &$\delta$ \\
 \hline MnO$_{2}$    &    0.235         &    0.170  &  --0.065\\
                            LaO   &    0.265      &    0.255 & --0.010   \\ \hline
\end{tabular}\end{table}

In addition,  the  MnO$_{2}$ surface plane is rumpled, as sketched in Fig.\ref{slab-sketch} and summarized in Table \ref{off-centering_table}, with Mn atoms being  strongly
displaced upwards relative to O atoms. The same pattern, even in quantitative terms, appears in the first  LaO layer from the surface. The rumpling is very large: the Mn-O (La-O) height difference is over two (over twenty) times the analogous  Ti-O (Pb-O) displacement in the typical ferroelectric\cite{fe} titanate PbTiO$_{3}$.
  This displacement generates  a  large surface dipole pointing out of the solid into the vacuum. In particular, the  FM surface has a larger near-surface rumpling  than  the  AF surface (Table \ref{off-centering_table}; effectively, the only significant structural difference between the two). This  is mirrored in a reduced workfunction of the FM  (W=4.47 eV for FM and W=4.67 eV  for AF). Of course, this matches the usual trend of metal work function lowering  with  increased surface corrugation.  Based on this result, a work function increase in a strained LSMO film would be a diagnostic evidence of a FM-to-AF magnetic phase change. 
  
\begin{figure}[h]
\includegraphics[width=7.5cm]{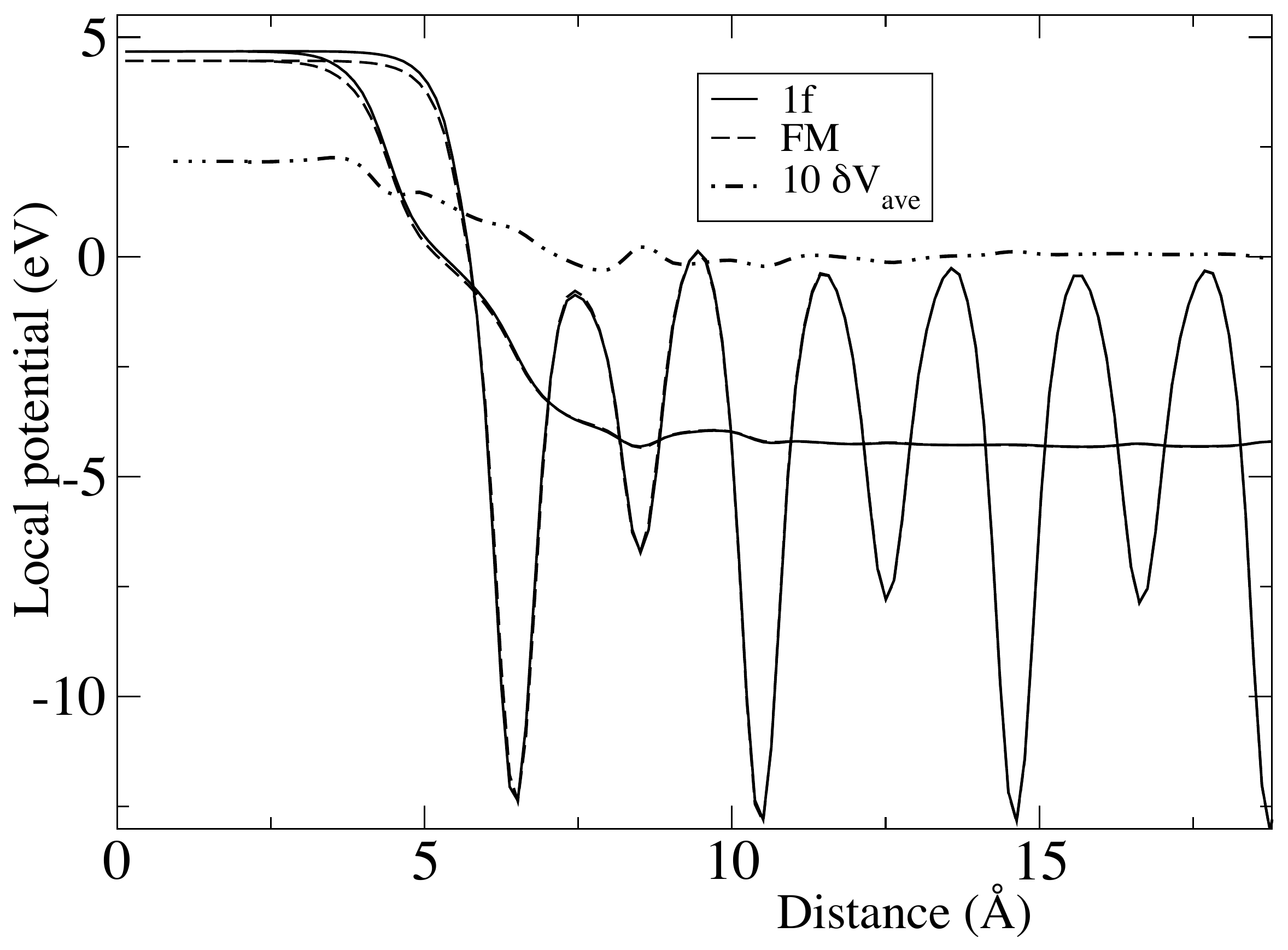}
\caption{(Color online) Local potentials and their periodicity-filtering running averages for  FM (dashed) and 1f surfaces (solid) between surface (left) and slab center (right). Energy zero is the Fermi energy. The dash-dot line at the center is the difference $\delta$V$_{\rm ave}$ of the 1f and FM averages (amplified by a factor 10).
\label{pot}}
\end{figure}

This argument is validated  by  Fig.\ref{pot}, which displays the local  potentials and their periodicity-filtering running averages for  the FM (dashed) and 1f surfaces (solid). The local potential has deep troughs at the atomic positions (the deepest ones are Mn, the others La), while the filtered averages are about constant in the bulk and in the vacuum. The dash-dot line at the center is the difference $\delta$V$_{\rm ave}$ of the 1f and FM averages amplified by a factor 10: clearly $\delta$V$_{\rm ave}$  is  zero in the bulk, it starts being perturbed by the surface at the second Mn layer, it  picks up decidedly from zero  just below the top Mn layer as expected from the dipole difference, it grows steadily as the surface is crossed, and finally saturates to the work function difference (0.2 eV) in the vacuum.

\begin{figure}[h]
\includegraphics[width=4.1cm]{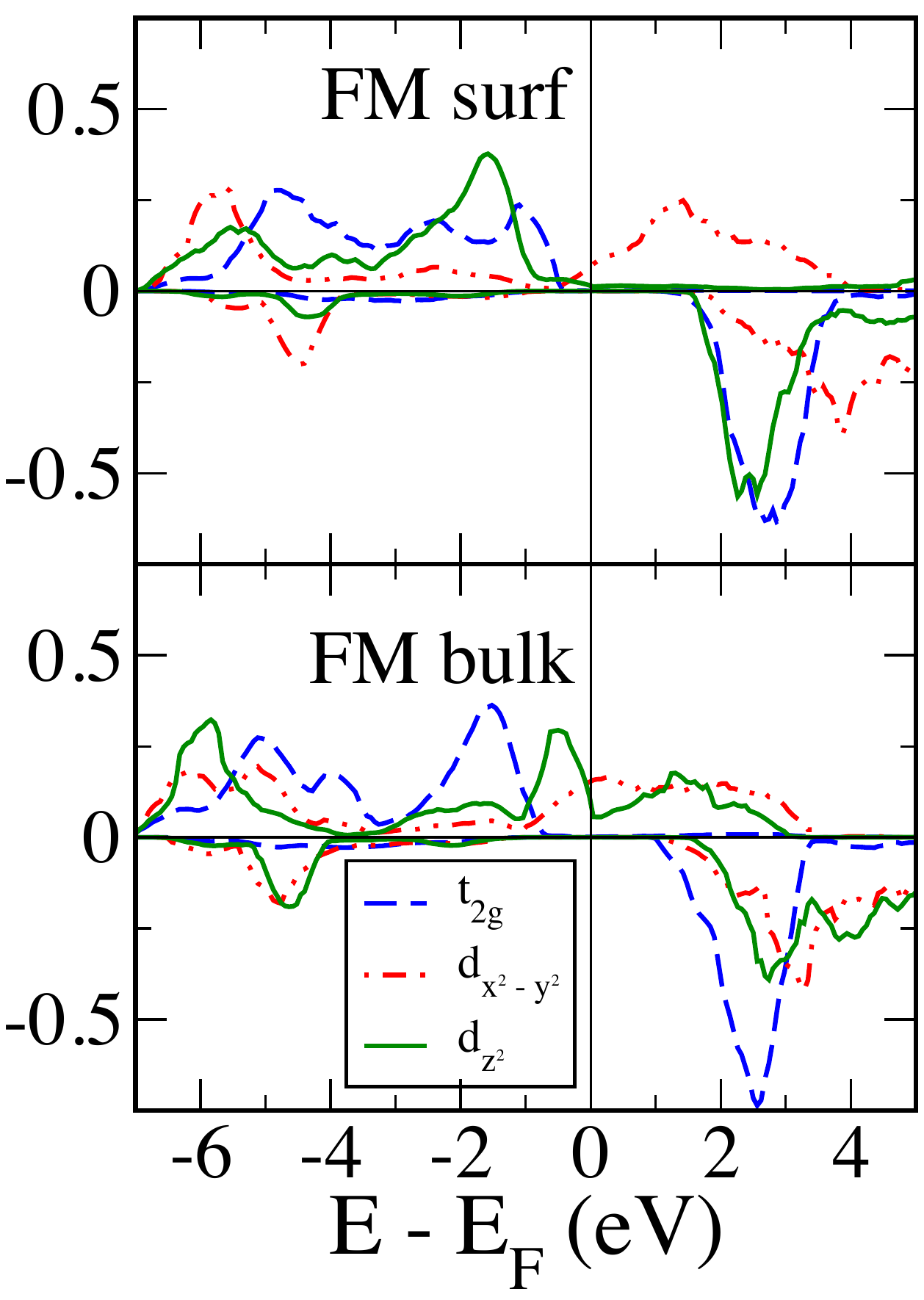}
\includegraphics[width=4.21cm]{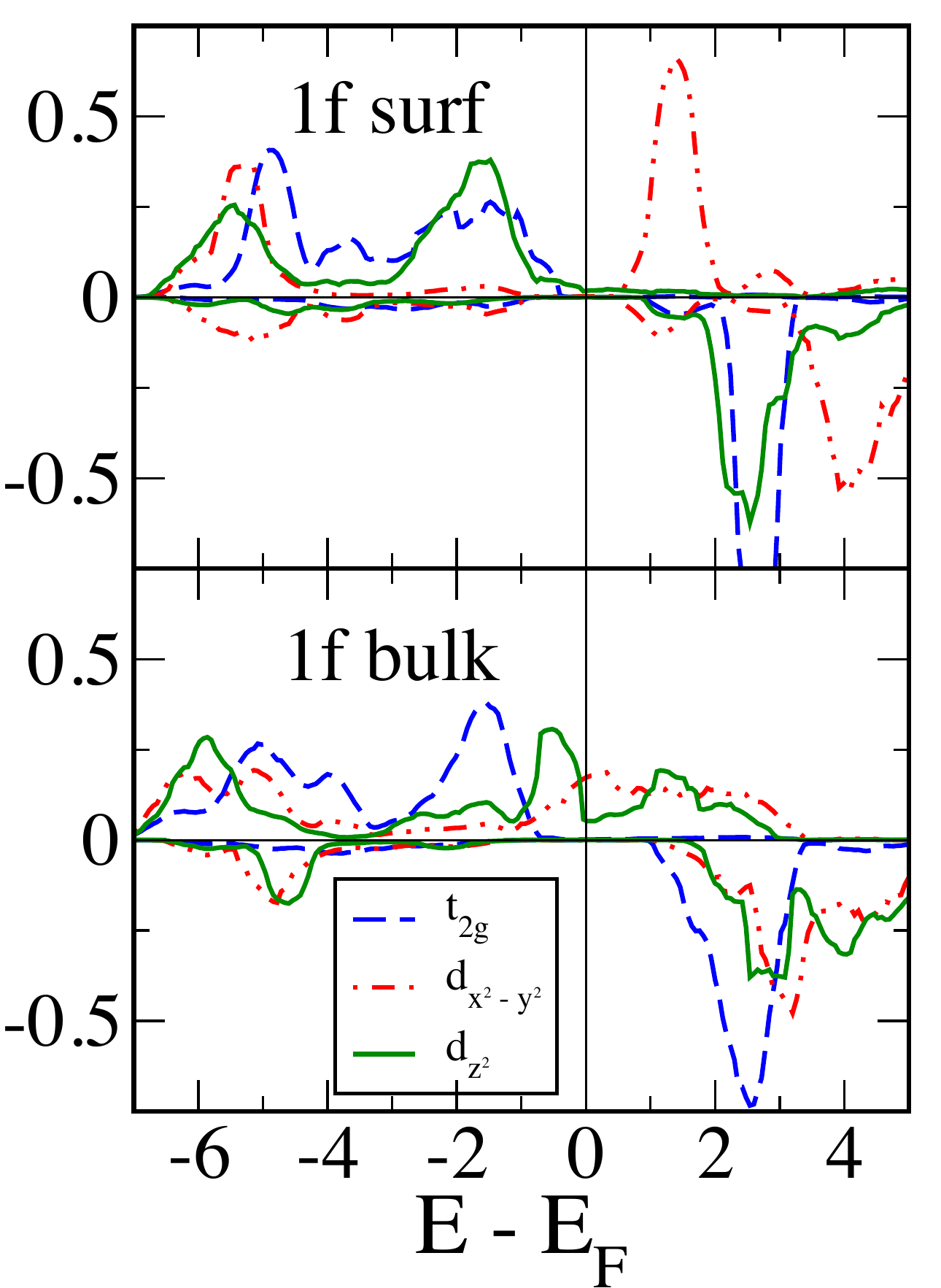}
\caption{(Color online) Orbitally resolved DOS for Mn at the surface (top) and in the innermost layer (bottom) in the FM (left) and 1f (right) magnetic order. Majority (minority) spin is positive (negative). Both surfaces are metallic, but the 1f has a much smaller DOS. The bulk DOS is identical in the two cases, demonstrating that the AF surface  does not affect bulk magnetism  beyond about 1 nm depth. 
\label{ordos-mn}}
\end{figure}

The  electronic structure  signature of this ``ferroelastic'' \cite{pruneda07} distortion is  a surface weakening of the directional $d_{z^{2}}$-mediated bonding, as borne out by Fig.\ref{ordos-mn},  discussed below. 
A similar rumpling of  MnO$_{2}$  planes (La-O layer rumplings were not mentioned) at FM, unstrained LSMO (001) was reported earlier  \cite{pruneda07} to decay gradually into the bulk. In the present case of  in-plane compression this feature is much stronger and  more surface-localized: our rumpling is twice as large (0.27 \AA\, vs 0.14 \AA), and it occurs only in the   outermost MnO$_2$ and LaO layers, while it is an order of magnitude lower, i.e. negligible, in the  other layers.

 Fig.\ref{ordos-mn} reports the  orbitally-resolved DOS projected on Mn for the FM (left) and 1f (right) magnetic order, at the surface (top) and in the innermost layer (bottom). The bulk DOS in the lower panels is identical in the two cases, demonstrating further that the AF surface  does not affect bulk magnetism  beyond at most 1 nm depth; also,  half-metallicity (a gap in the minority channel) is preserved in all cases both in the bulk and at the surface. While the bulk is metallic with roughly equal participation of both $e_{g}$ orbitals, the FM surface is still fully metallic, but with mostly $d_{x^{2}-y^{2}}$ and only residual $d_{z^{2}}$ character  at E$_{F}$. The standard DOS narrowing and upward shift at the surface (which would  provide strong $d_{z^{2}}$  metallicity) is prevented by the huge outward relaxation, which indeed is associated with the $d_{z^{2}}$ peak dipping below E$_{F}$. The  AF 1f surface has a real gap in both the $t_{2g}$ and $d_{x^{2}-y^{2}}$ channels. On the other hand,  the $d_{z^{2}}$ weight moves below E$_{F}$ as in the FM, but this demotion is incomplete and a small but appreciable DOS survives at E$_{F}$. We thus end up with a  deep surface-related pseudogap, signaling an incipient but incomplete metal-insulator transition at the surface. This  is still technologically relevant, since  conduction through a contact on this surface in, e.g., a tunneling-magnetoresistance or  field-effect device would clearly be affected by this 1-eV-wide pseudogap.

\begin{figure}[h]

\includegraphics[width=7.5cm]{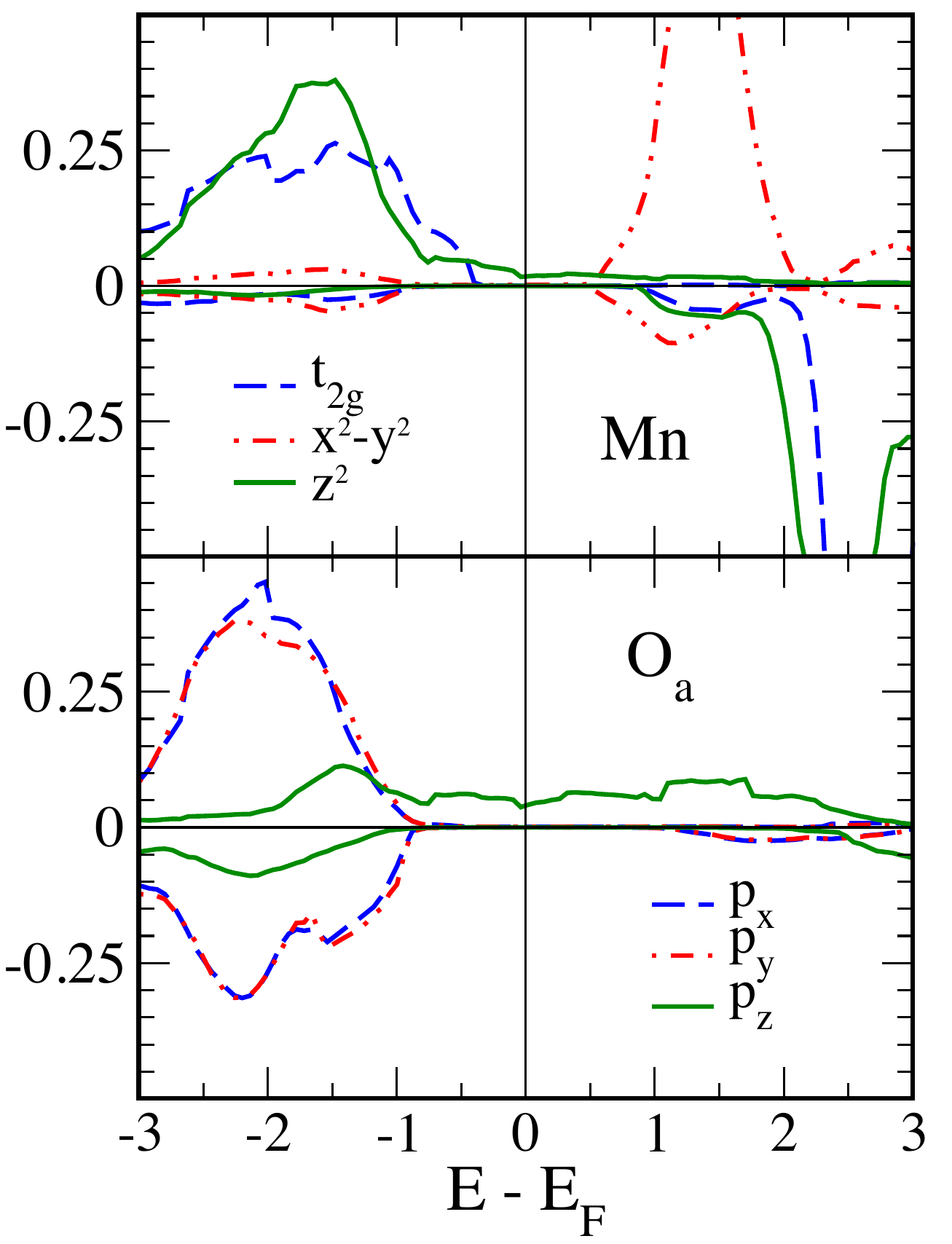}
\caption{(Color online) A close-up view in the Fermi level region of the DOS projected on the spin-up surface Mn  and the corresponding apical subsurface O  for the  1f structure (see also Fig.\protect\ref{ordos-mn}, top right). The residual metallic character  is due to the vertical, non-bonding  overlap of  Mn $d_{z^{2}}$ with O$_{a}$ $p_{z}$. 
\label{ordos-mn2}}	
\end{figure}

 The marginal (in the sense of being related to a pseudogap,  for example as in bulk Be) character  of  surface metallicity in the surface AF phase is further illuminated by  the orbitally resolved DOS for the spin-up Mn in the AF top layer  and the corresponding  apical oxygen O$_{a}$ in the subsurface LaO layer, displayed in Fig.\ref{ordos-mn2}, bottom.   Clearly, the residual metallic character  is due to the residual (non-bonding)  overlap of  Mn $d_{z^{2}}$ with O$_{a}$ $p_{z}$ in the majority channel (solid lines), which confirms that the in-plane orbital channels are all gapped.  The spin-down Mn and corresponding O$_{a}$ (not shown) behave similarly, with a non-zero DOS in the majority channel, that preserves the (by now weak) semimetallic character.
Despite the absence of a proper gap, it is  likely that the conductivity will be strongly diminished, especially in-plane, due to the drop in the DOS at E$_{F}$. 

In closing, we note that the surface electronic structure just discussed is fully consistent with the strain effects on magnetic coupling mechanisms discussed in Ref.\onlinecite{colizzi-PRB08} (see in particular the left panel of Fig.14 thereof): basically, in-plane compression favors rotations around the vertical axis $\hat{s}$, and unbalances the magnetic couplings so that double-exchange (FM) tends to prevail along $\hat{s}$ while super-exchange (AF) is stronger in-plane (also, a charge transfer from $d_{x^{2}-y^{2}}$ to $d_{z^{2}}$ attends to this shift). However, very large  strains would be needed in the bulk for   AF to prevail; the surface provides locally the additional structural freedom to (barely) tip the balance in AF's favor in some strain range, realizing locally a forerunner of magnetic phase separation.\cite{tebano-PRB06}
 
In summary, we have shown that under compressive epitaxial strain  the  first  Mn surface layer  (possibly, at most, the first two) of ferromagnetic LSMO (001) turns antiferromagnetic. This change in magnetic order is accompanied by a  marked surface-related spectral pseudogap, signaling an incomplete metal-insulator transition at the surface.
The LSMO surface is under significant tensile stress, which favors accommodation on compressing epitaxial substrates.
The  MnO$_{2}$ and LaO surface planes are severely  rumpled due to weakened  Mn $d_{z^{2}}$-O $p_{z}$ bonding, and the different rumpling in the two magnetic phases causes a distinct  work function shift of potential diagnostic use. The surface AF phase  is an extreme surface-assisted case of a previously recognized \cite{colizzi-PRB08} combination of in-plane AF super-exchange and vertical FM double-exchange couplings under in-plane compression. 

Work supported in part by projects EU FP7 {\it OxIDes} (grant n.228989),  MIUR-PRIN 2008 {\it 2-DEG FOXI} and 2010 {\it Oxide}, IIT Seed {\it NEWDFESCM}, Fondazione Banco di Sardegna 2011. Computational resources provided by CASPUR Rome and by KAUST HPC.

\end{document}